\documentclass[epj]{svjour}
\usepackage{latexsym}
\usepackage{graphicx}
\usepackage{amsmath}
\begin{document}
\title{Magnetic properties of a family of quinternary oxalates}
\author{E. Lhotel \inst{1} \and V. Simonet \inst{1} \and J. Ortloff \inst{1,2}\and B. Canals \inst{1} \and C. Paulsen \inst{1} \and E. Suard \inst{3} \and T. Hansen \inst{3} \and D. J. Price \inst{4} \and P. T. Wood \inst{5} \and A. K. Powell \inst{6,7} \and R. Ballou \inst{1}}
\mail{elsa.lhotel@grenoble.cnrs.fr} 
\institute{Institut N\'eel, CNRS \& Universit\'e Joseph Fourier, BP 166, 38042 Grenoble Cedex 9, France \and Institute for Theoretical Physics, University of W\"urzburg, Germany \and Institut Laue Langevin, BP 156, 38042 Grenoble Cedex 9, France \and Univ Glasgow, School of Chemistry, Glasgow G12 8QQ, Scotland \and Univ Cambridge, Chem Lab, Cambridge CB2 1EW, England \and Institute of Inorganic Chemistry, Karlsruhe Institute of Technology, Engesserstrasse 15, D-76131 Karlsruhe, Germany \and Institute for Nanotechnologie, Karlsruhe Institute of Technology, Postfach 3640, D-76021 Karlsruhe, Germany }
\date{Received: date / Revised version: date}
%
\abstract{We report on the magnetic properties of four isomorphous compounds of a family of quinternary oxalates down to 60 mK. In all these materials, the magnetic Fe$^{\rm II}$ ions with a strong magneto-crystalline anisotropy form a distorted kagome lattice, topologically equivalent to a perfect kagome one if nearest-neighbor interactions only are considered. All the compounds order at low temperature in an antiferromagnetic arrangement with magnetic moments at 120$^{\circ}$. A remarkable magnetic behavior emerges below the N\'eel temperature in three compounds (with inter-kagome-layer Zr, Sn, Fe but not with Al): the spin anisotropy combined with a low exchange path network connectivity lead to domain walls intersecting the kagome planes through strings of free spins. These produce an unfamiliar slow spin dynamics in the ordered phase observed by AC susceptibility, evolving from exchange-released spin-flips towards a cooperative behavior on decreasing the temperature. 
\PACS{
      {75.25.-j}{Spin arrangements in magnetically ordered materials}   \and
      {75.60.Ch}{Domain walls and domain structure} \and
      {75.40.Gb}{Dynamic properties} \and
      {75.50.Ee}{Antiferromagnetics}
       } 
} 

\maketitle
\section{Introduction}
\label{intro}

When induced by the topology of the lattice, magnetic frustration is expected to produce exotic phases and highly degenerate ground states with short-range spin-spin correlations \cite{Moessner2006}. These are characterized for instance by a 120$^{\circ}$ spin arrangement on each triangle of a classical two-dimensional corner-sharing triangle kagome lattice with antiferromagnetic nearest-neighbor (NN) interactions. The predicted resulting spin liquid state in the Heisenberg kagome antiferromagnet has still to be found unambiguously in real materials where it is in practice easily destabilized in presence of additional parameters beyond the NN interactions (next nearest neighbors (NNN) exchange interaction, single-ion anisotropy, dipolar interaction, Dzyaloshinsky-Moriya antisymmetric exchange interaction, non-stoichiometry...). These extra terms in the hamiltonian may nevertheless lead to unexpected new physics as is the case for the spin-ices discovered in the pyrochlore materials \cite{Gingras11}. There, dipolar interactions associated to a strong multiaxial anisotropy (along the [111] direction) lead to a degenerate ground state constrained only by the two-in two-out spin configuration on each tetrahedron.
Magnetic frustration does not only engender new ground states but also associated remarkable dynamics. For instance, propagative spin-wave like modes were calculated to emerge from the disordered spin liquid state of the Heisenberg kagome model \cite{Robert08} as well as weathervane soft local modes \cite{Ritchey93,Keren94}. Another example are the excitations supported by the dipolar spin ices which behave like deconfined magnetic monopoles and give rise to slow dynamics \cite{Ryzhkin05,Castelnovo08,Jaubert09}. 

Thanks to its versatility, molecular chemistry is a promising way to obtain new frustrated materials, such as kagome systems. In this search for new kagome compounds, oxalate ligands have little been exploited, although they are known to stabilize rather strong magnetic interactions, and have been used previously in the synthesis of honeycomb systems \cite{Tamaki92}, which are potentially frustrated beyond NN interactions. 
Recently, new molecular kagome compounds with other ligands have been synthesized \cite{Awaga94,Liu99,Aidoudi11}. Due to the small value of the spins involved, they are all considered as realizations of a quantum kagome lattice. 

In this article, we report our magnetic measurements on a family of quinternary oxalates, which is a realization of a classical kagome like lattice. We extend the results presented in reference \cite{Lhotel11} by studying four members of this family. After a brief presentation of the structure of the compounds, we study their magnetic behavior in the paramagnetic regime. Then we show that they order in a 120$^{\circ}$ antiferrromagnetic structure at temperatures of a few Kelvin. Finally, we discuss their dynamic behavior below the transition temperature, and show that in three of the studied compounds unconventional spin dynamics occur within domain-walls.

\section{The quinternary oxalates family}
\label{sec1}

Two new isostructural series of kagome antiferromagnets were synthesized using hydrothermal methods: series I with the composition Na$_2$Ba$_3$[Fe$^{\rm II}_3$(C$_2$O$_4$)$_6$][A$^{\rm IV}$(C$_2$O$_4$)$_3$] where A$^{\rm IV}$ = Sn$^{\rm IV}$, Zr$^{\rm IV}$; and series II with the composition Na$_{2}$Ba$_3$ [Fe$^{\rm II}_3$(C$_2$O$_4$)$_{6}$][A$^{\rm III}$(C$_2$O$_4$)$_3$]$_{0.5}$[A$^{\rm III}$(C$_2$O$_4$)$_2$(H$_2$O)$_2$]$_{0.5}$, where A$^{\rm III}$ = Al$^{\rm III}$, Fe$^{\rm III}$. 
In the following, these oxalate compounds will be abbreviated QO-FeA referring to the common divalent Fe$^{\rm II}$ and the tri- or tetravalent ion A=Sn, Zr, Al, Fe. 

\begin{figure}
\begin{center}
\includegraphics[width=7cm]{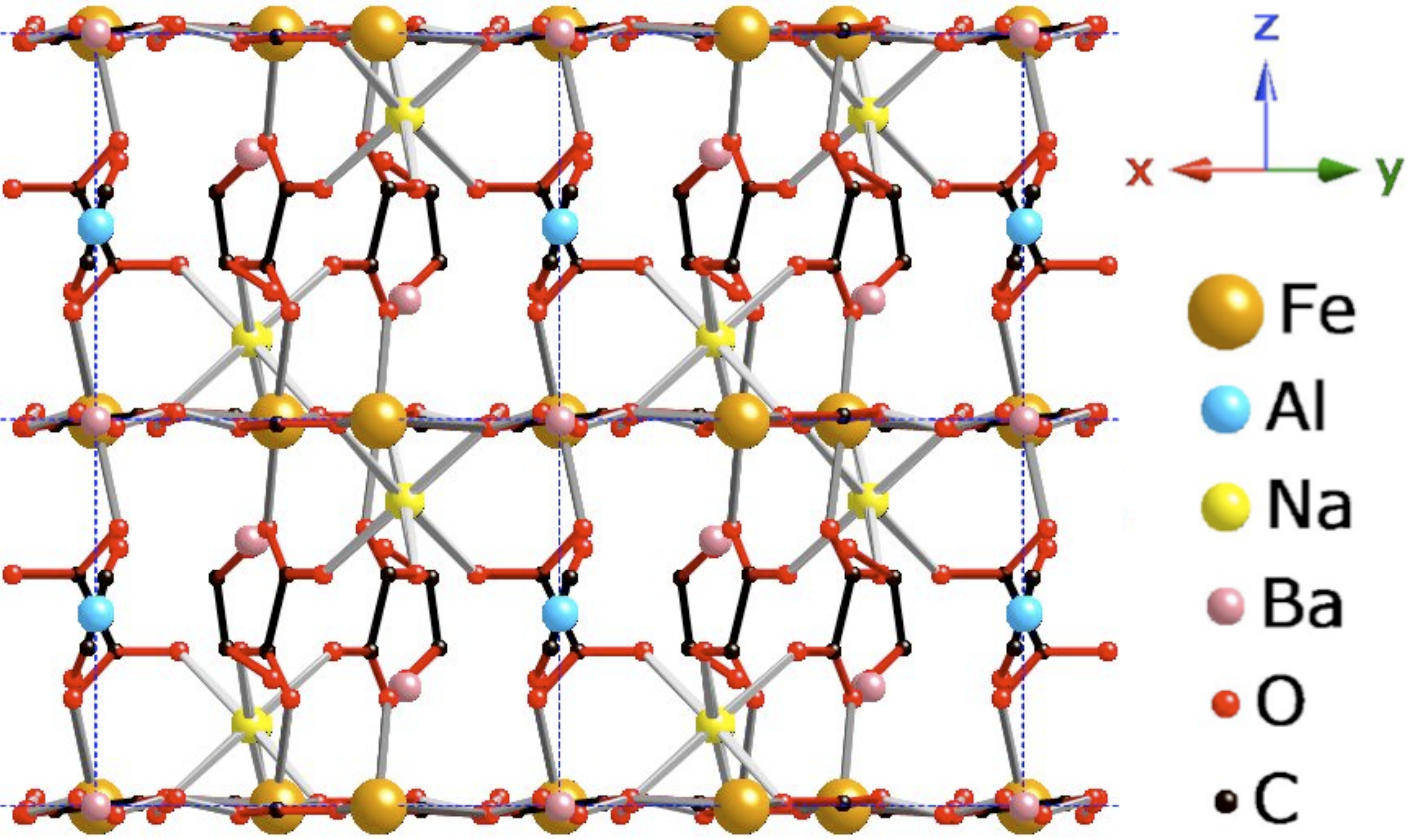}
\includegraphics[width=7cm]{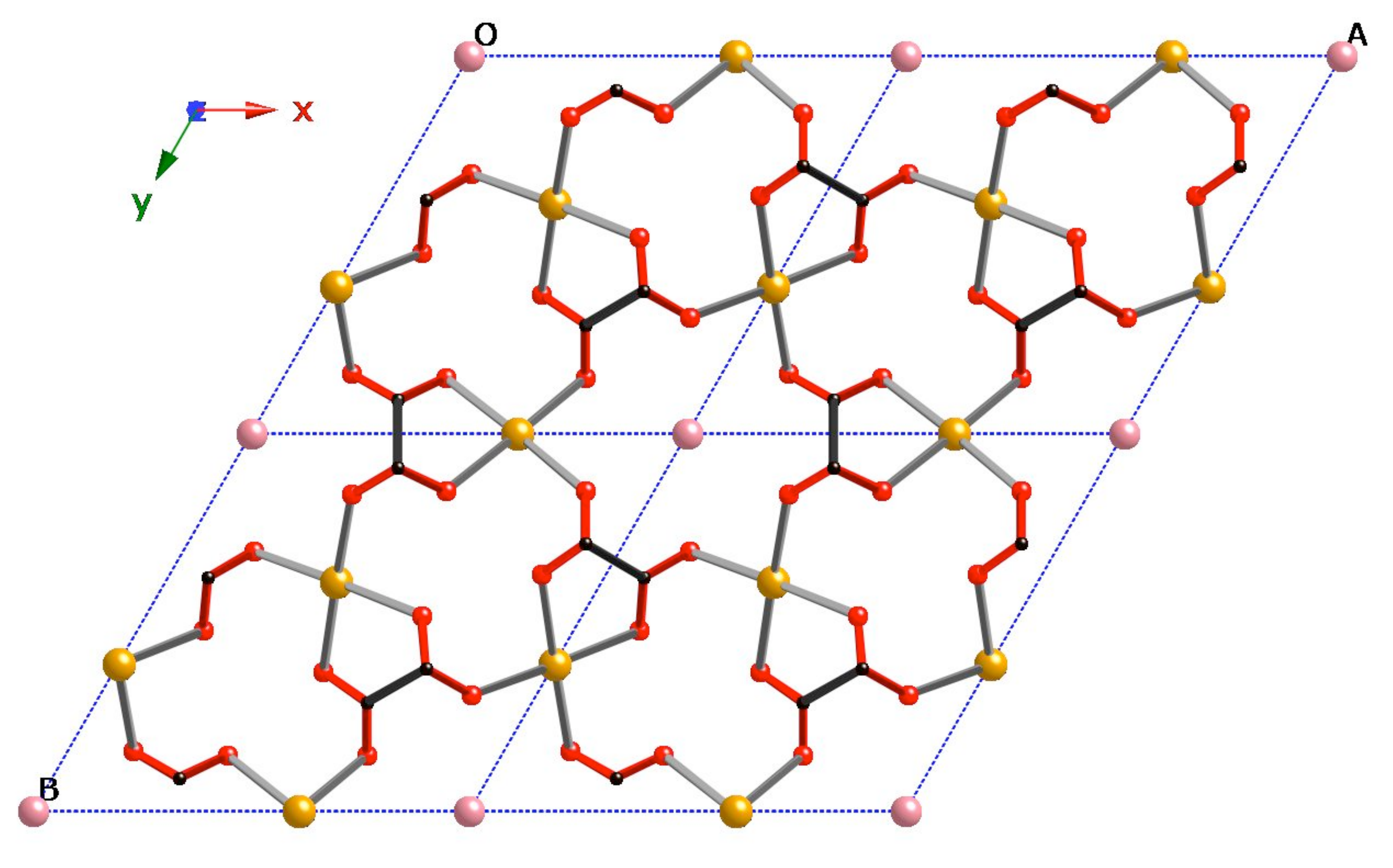}
\end{center}
\caption{Structure of QO-FeA series II on a plane perpendicular to the ${\bf a}+{\bf b}$ direction (top) and on the (${\bf a}$, ${\bf b}$) plane (bottom) with $a=b=10.45$ \AA, $c=7.54$ \AA. There are three equivalent Fe$^{\rm II}$ per unit cell on the Wyckoff site 3$e$ at positions (0, 0.6145, 0), (0.6145, 0, 0) and (0.3854, 0.3854, 0).} 
\label{fig_ox}
\end{figure}

All the compounds are isomorphous. They crystallize in the non-centrosymmetric trigonal P321 space group with $a=b=10.45$ \AA, $c=7.54$ \AA . In both series, the [Fe$^{\rm II}_3$(C$_2$O$_4$)$_6$] network forms, in the ($a$, $b$) plane, a distorted kagome lattice of Fe$^{\rm II}$, which are stacked along the $c$ axis. The distortion from an ideal kagome lattice is a reduction of the lattice symmetry from six to three-fold. However if we consider only NN interactions, the Fe$^{\rm II}$ lattice is topologically equivalent to the kagome one (see Fig. \ref{fig_ox} bottom). This equivalence is broken if we consider NNN interactions. The remaining ions in the structure occupy the inter-layer space (see Fig. \ref{fig_ox} top). 

The magnetic exchange interactions are mediated via oxalate C$_2$O$_4^{2-}$ dianions. Intraplane NN and NNN interactions, called $J_1$ and $J_2$ respectively, involve the same oxalate bridge, in a mode that has not been reported before to our knowledge (see Fig. \ref{fig_mag}), thus making the estimate of the exchange interactions rather complicated. Similar first neighbor paths were found to be negligible in some Fe oxalate compounds \cite{Mennerich08} but were evaluated between 0.5 and 2 K in Cu oxalate compounds \cite{Kikkawa05,Nunez01}. Similar kinf of bridging as the NNN path was also reported in reference \cite{Price01}, but in the latter case, four Fe ions were involved. The inter-plane exchange path $J_3$ on the other hand is mediated via an oxalate ligand in a {\it trans} mode as in reference \cite{Fei05}. The $J_2$ and $J_3$ exchange interactions are anyway expected to be much weaker than $J_1$ since the exchange path is longer ($l_2\approx 7$~\AA~compared to $l_1=5.591$ \AA) and involves two C of the bridging oxalate.

\section{Experimental details}
We measured the powder sample magnetization and AC susceptibility of the four compounds by the extraction method, using a Quantum Design MPMS magnetometer for temperatures above 2 K and a superconducting quantum interference device magnetometer equipped with a miniature dilution refrigerator developed at the Institut N\'eel for temperatures down to 70 mK \cite{Paulsen01}. Neutron diffraction measurements were performed on the two two-axis diffractometers D20 and D2B with a wavelength equal to 2.4 \AA\ at the Institut-Laue-Langevin high-flux reactor. Diffractograms were recorded in a cryostat down to 1.5~K on the four compounds (deuterated for series II), and using a dilution insert down to 60~mK on the QO-FeZr compound. 

\section{Magnetic behavior above 2 K}
\label{sec2}
\subsection{"High" temperature magnetic properties}

\begin{figure}
\begin{center}
\includegraphics[width=7.5cm]{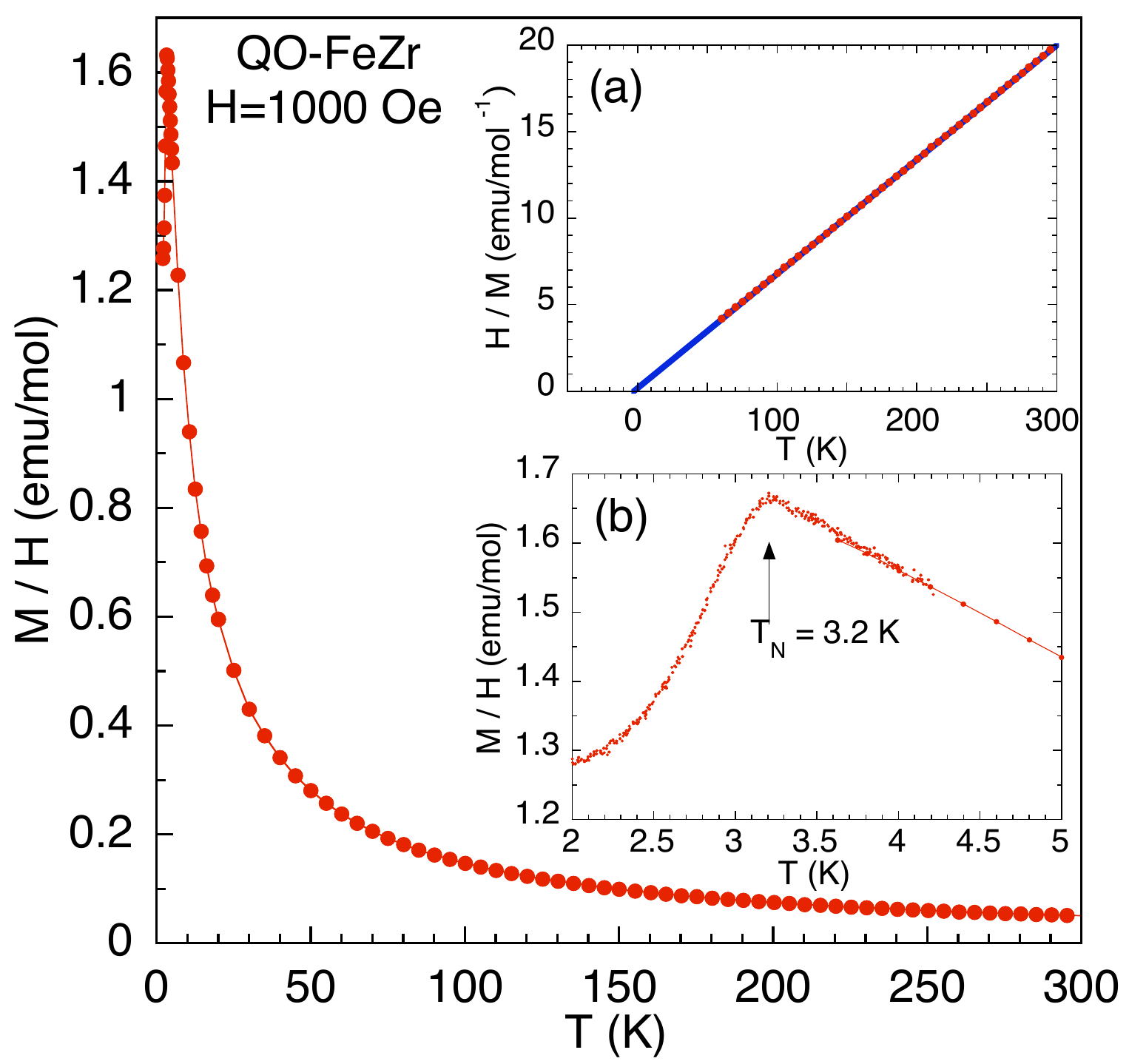}
\end{center}
\caption{$M/H$ vs $T$ for QO-FeZr in a 1000 Oe magnetic field. Insets: (a) $H/M$ vs $T$. The line is a fit using the Curie-Weiss law with $\theta=-5.1$ K and $C=15.5$ emu.mol$^{-1}$.K$^{-1}$. (b) Zoom of $M/H$ vs. $T$ around the N\'eel temperature.} 
\label{fig_FeZr_MT}
\end{figure}

In the high temperature linear regime, the susceptibility $\chi=M/H$ of all the compounds is accounted for down to 50 K by a Curie-Weiss model  $C/(T-\theta)$ where $C={\cal N}_A\mu_{\rm eff}^2 / 3k_B$ with the effective moment $\mu_{\rm eff}=g_J\mu_B\sqrt{J(J+1)}$ (see inset (a) of Fig. \ref{fig_FeZr_MT}). The only magnetic ions are the Fe$^{\rm II}$ ions except in QO-FeFe where paramagnetic Fe$^{\rm III}$ ions present between the kagome planes also contribute to the magnetization. The analysis of the Curie-Weiss constant extracted from the susceptibility and of the magnetization versus field allowed to establish that the Fe$^{\rm III}$ are in the low spin state ($S=1/2$). This surprisingly is at variance with the high spin value ($S=5/2$) of reference \cite{Mathoniere96} where the Fe$^{\rm III} $ ions are similarly octahedrally coordinated. It suggests a proximity to a low spin - high spin instability, not observed in the temperature range of our measurements. In QO-FeFe, the  Fe$^{\rm III}$ spins are without significant interactions down to very low temperature. 
After subtracting their contribution in order to isolate the Fe$^{\rm II}$ behavior in the QO-FeFe, the Curie-Weiss temperature $\theta$ for all the compounds ranges from -5 to -9 K and $\mu_{\rm eff}$ varies between 6 and 7 $\mu_B$ (see Table \ref{table}), which is not significantly different within the error bars. The effective moments thus obtained are close to the value of 6.7 $\mu_B$ expected for the free ion with $J=L+S=4$ ($L=2$ and $S=2$). 

\begin{table}
\caption{Summary of the main parameters characterizing the magnetic behavior above 2 K of the QO-FeA compounds. $M_{\rm sat}$ is the value of the magnetization at 8 T and 70 mK. The estimated maximum error bars of the Curie-Weiss fit are 2 K for $\theta$ and 0.3 for $\mu_{\rm eff}$.}
\label{table}
\begin{tabular}{|*{5}{c|}}
\hline\noalign{\smallskip}
A & $T_{\rm N}$ (K) & $M_{\rm sat}$ ($\mu_{\rm B}$/Fe$^{\rm II}$)& $\theta$ (K) & $\mu_{\rm eff}$  ($\mu_{\rm B}$/Fe$^{\rm II}$)\\
\noalign{\smallskip}\hline\noalign{\smallskip}
Zr & 3.2 & 3.8 & -5.1 & 6.4 \\
Sn & 3.2 &3.4 & -9.4 & 7.1 \\
Fe & 3.2 & 3.8 & -8.6 & 6.5 \\
Al & 1.9 & 3.1 & -9.2 & 6 \\
\noalign{\smallskip}\hline
\end{tabular}
\end{table}

\begin{figure}
\begin{center}
\includegraphics[width=8cm]{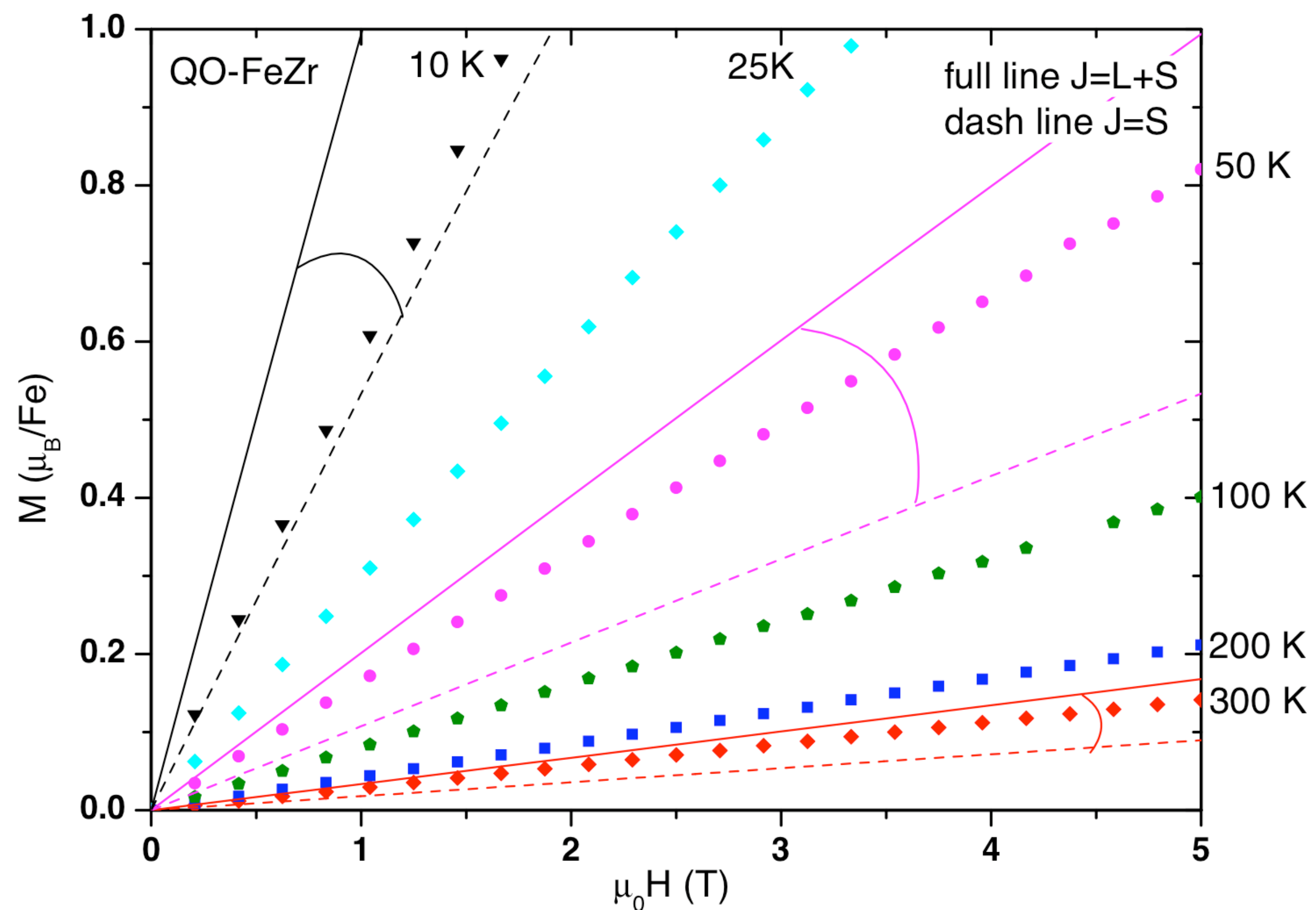}
\end{center}
\caption{$M$ vs $H$ between 10 K and 300 K for QO-FeZr. The full lines (respectively the dashed lines) correspond to a Brillouin law with $J=L+S=4$ (respectively $J=S=2$ with $L$=0), at 10, 50 and 300 K. } 
\label{fig_MH_HT}
\end{figure}

At 300 K, the magnetization isotherms are indeed close to a Brillouin curve calculated with a total angular momentum $J=4$ (see Fig. \ref{fig_MH_HT}). However, below 100 K, magnetization isotherms gradually depart from the $J=4$ Brillouin curves,  and at low temperature, below 5~K, the saturation magnetization in an applied field of 10 T is nearly 4 $\mu_B$. Structural patterns recorded by neutron diffraction show that these features are not due to a structural distorsion associated with a Jahn-Teller like transition as proposed in reference \cite{Fishman}. 
The low value of the saturation magnetization might instead be due to the magneto-crystalline anisotropy, estimated around ten Kelvin, as detailed in next Section.

\subsection{Magnetic structure below T$_N$}

\begin{figure}
\begin{center}
\includegraphics[width=7.5cm]{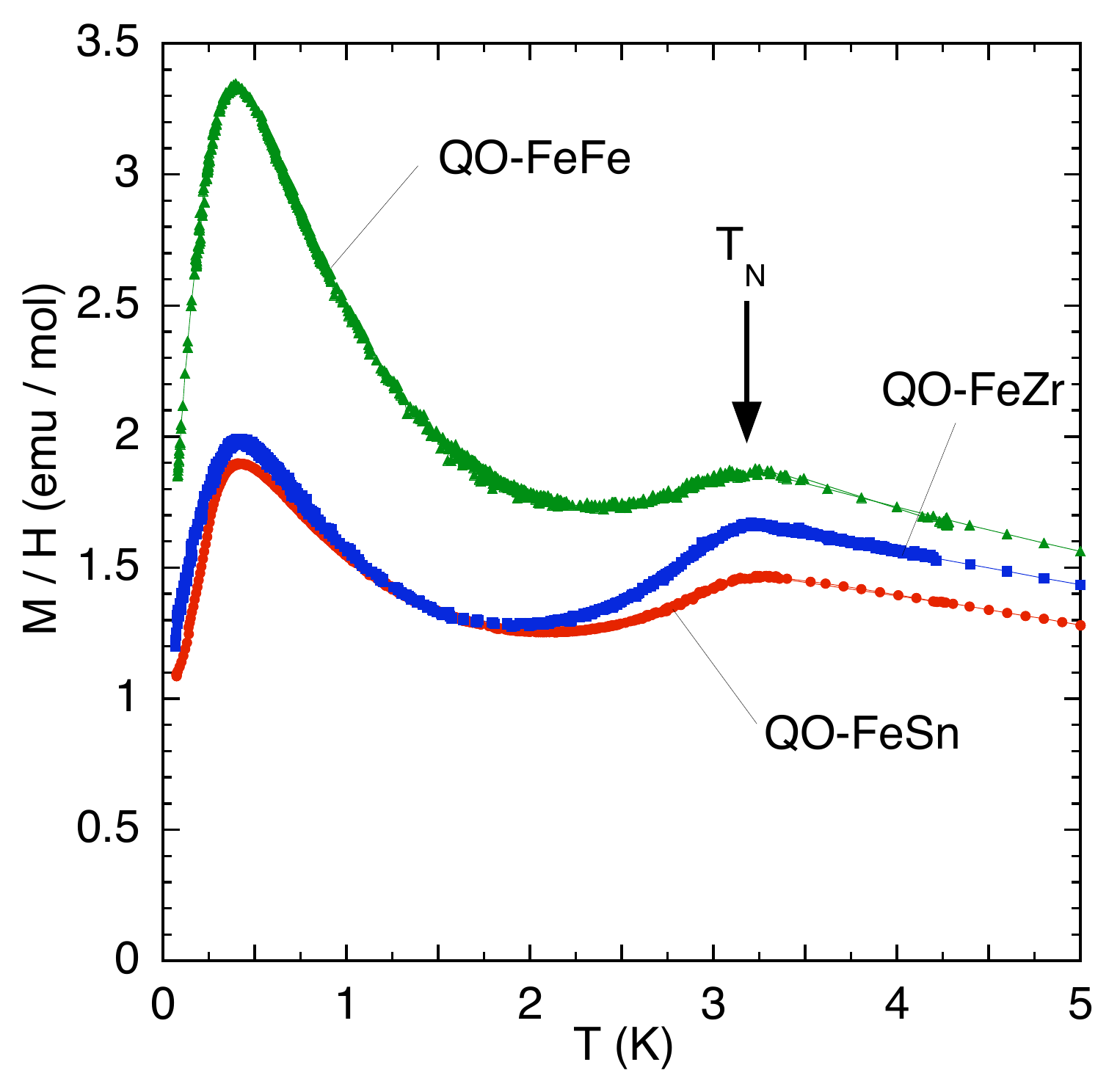}
\end{center}
\caption{$M/H$ vs $T$ for QO-FeA with A=Fe, Zr, Sn in a 500 Oe applied magnetic field. The increase of the magnetization when decreasing the temperature is larger in the QO-FeFe compound, which is due to the paramagnetic contribution of the spins 1/2 of the Fe$^{\rm III}$ ions. } 
\label{fig_ox_MT}
\end{figure}

At low temperature, a cusp characteristic of an antiferromagnetic transition is observed in the magnetization as a function of temperature (see inset (b) of Fig. \ref{fig_FeZr_MT}): the N\'eel temperature, equal to 3.2 K, is identical for the QO-FeA compounds with A=Zr, Sn, Fe (see Fig. \ref{fig_ox_MT}). For the QO-FeAl compound, it is slightly smaller with a value of 1.9 K (see Fig. \ref{fig_FeAl_MT}). The reasons for the difference between the Al based compound and the others will be discussed in Section \ref{discussion}. The obtained N\'eel temperatures $T_N$ are of the same order of magnitude as the Curie-Weiss temperature $\theta=-5$ K. This indicates a release of the high degeneracy expected for antiferromagnetic NN interactions in this kagome like system. 

\begin{figure}
\begin{center}
\includegraphics[width=8.5cm]{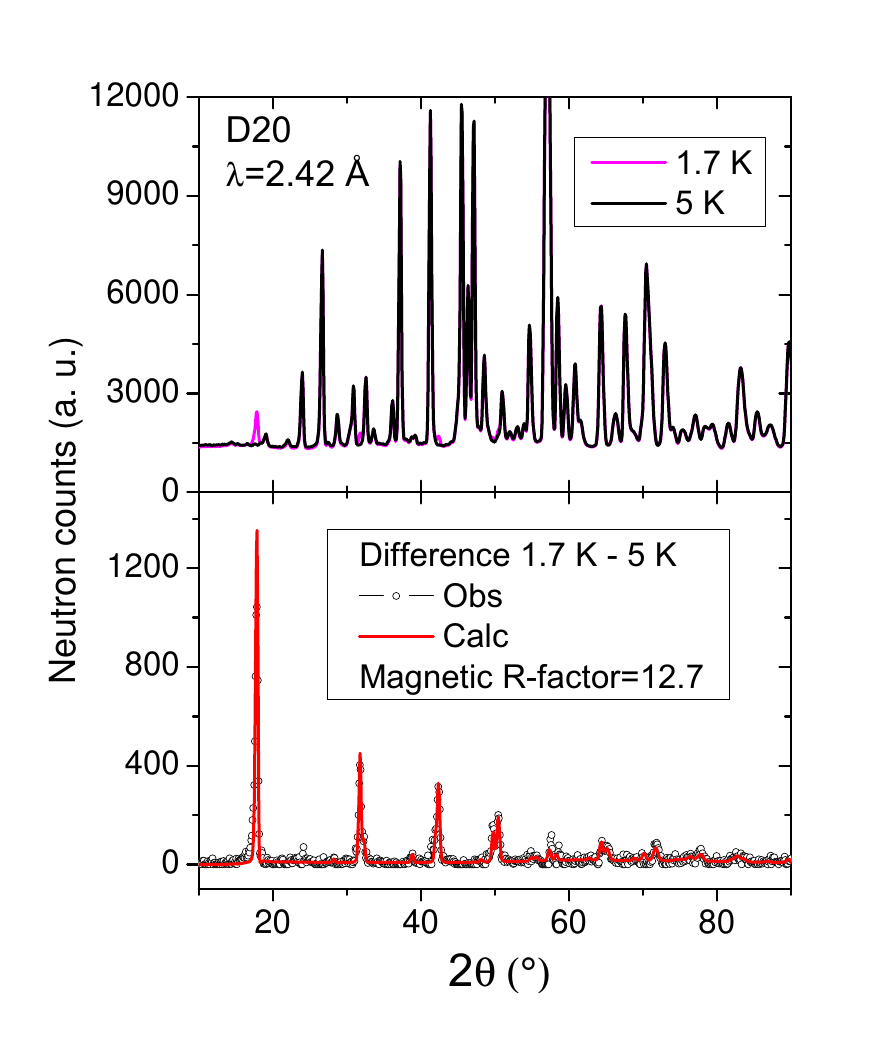}
\end{center}
\caption{(top) Neutron diffractograms of QO-FeFe above (5 K) and below (1.7 K) T$_N$ recorded on D20. (bottom) Magnetic neutron pattern obtained from the difference between the 1.7 and 5 K diffractograms (empty dot). It is compared to the refinement (red line) with a propagation vector (0, 0, 1/2) and the {\bf q=0} magnetic arrangement with magnetic moments at 120$^{\circ}$ in the (${\bf a}$, ${\bf b}$) planes, discussed in the text. } 
\label{fig_neut}
\end{figure}

\begin{figure}
\begin{center}
\includegraphics[width=7.5cm]{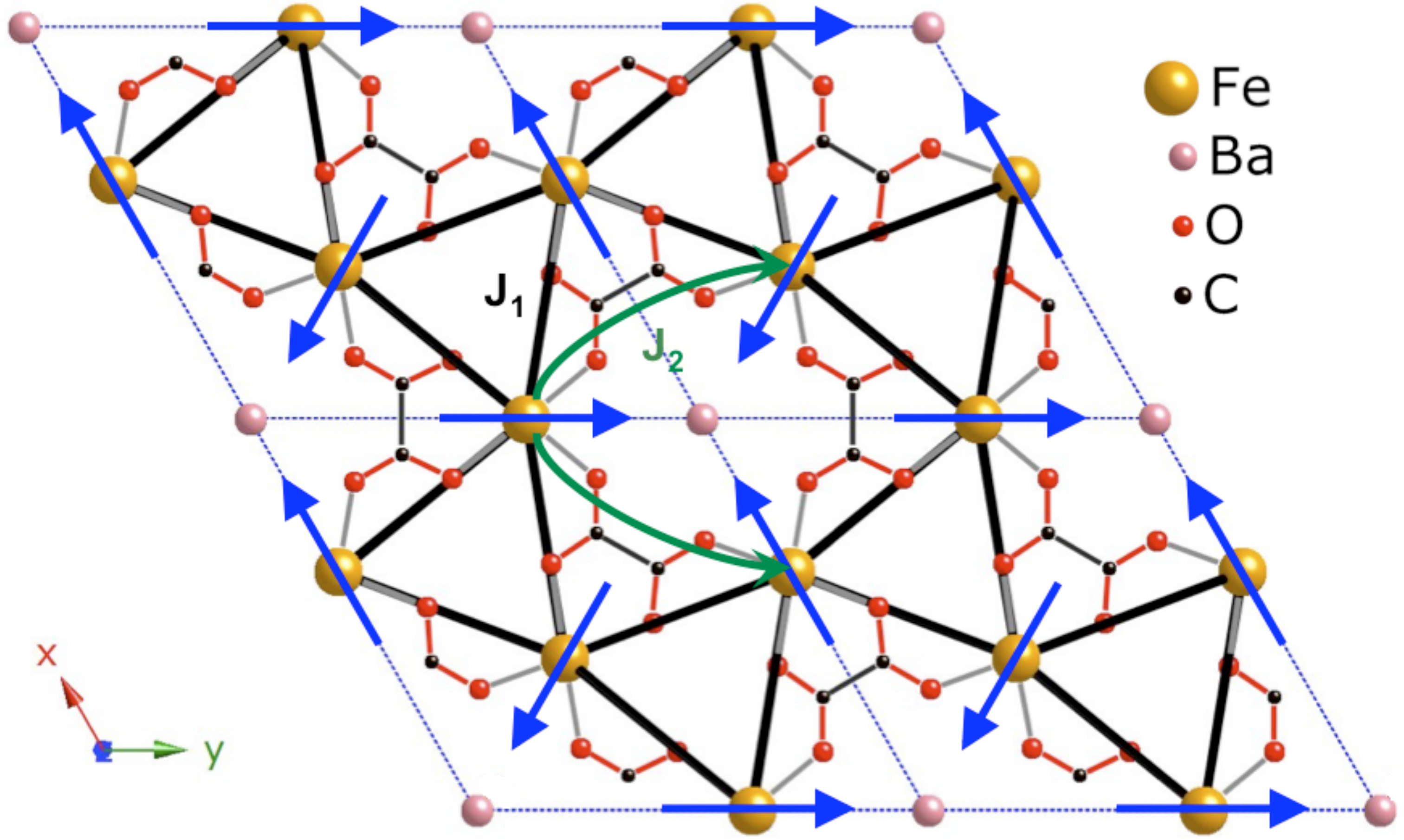}
\end{center}
\caption{Magnetic structure in the (${\bf a}$, ${\bf b}$) plane. The black lines materialize the Fe$^{\rm II}$ NN exchange interaction lattice. The NNN $J_2$ exchange interactions are shown by the green curved arrows. The blue arrows represent the ordered magnetic moments. } 
\label{fig_mag}
\end{figure}

Neutron diffraction measurements confirm the onset of a magnetic order, identical in all compounds. Additional reflections indeed rise below T$_N$ that can be indexed with a propagation vector ${\bf k}=(0,0,1/2)$ indicating an antiferromagnetic stacking along the $c$ axis (see Fig. \ref{fig_neut} for QO-FeFe and Fig. 3 of reference \cite{Lhotel11} for QO-FeZr). The magnetic order was refined from the difference between the 1.5 and 5 K diffractograms. The so-called {\bf q}=0 in-plane magnetic arrangement consists in nearest neighbor magnetic moments orientated at 120$^{\circ}$ from each other and lying along the ${\bf a}$, ${\bf b}$ and $-{\bf a}-{\bf b}$ axes, with the same chirality of spins for all the triangles (see Fig. \ref{fig_mag}). It corresponds to one of the three irreducible representations compatible with the crystal symmetries and leaving the propagation vector invariant.

Since no long range order is predicted for the classical kagome antiferromagnet with isotropic nearest-neighbor exchange interactions, additional mechanisms must be invoked to explain this ordered ground state. Note that a {\bf q=0} order with a 120$^{\circ}$ spin arrangement has also been reported in the Fe$^{\rm III}$ KFe$_3$(OH)$_6$(SO$_4$) jarosite \cite{Wills00}. The latter magnetic arrangement however differs from the present one: the 120$^{\circ}$ spins in the kagome planes are orientated perpendicular to the ${\bf a}$, ${\bf b}$ and $-{\bf a}-{\bf b}$ axes and an out-of plane component is present. From a spin-waves analysis \cite{Matan06}, it was explained by a Dzyaloshinsky-Moriya interaction preferred over a single-ion anisotropy, in addition to NN antiferromagnetic interactions. 

In the QO-FeA compounds, we can infer from the Curie-Weiss temperature $\theta$ that the exchange energy due to the NN interactions $J_1$ is between 3 and 5 K, which is consistent with reported values in the literature \cite{Mennerich08,Kikkawa05,Nunez01}. Comparison with other compounds involving Fe$^{\rm II}$ ions and oxalate ligands also suggests that there is a strong single-ion anisotropy of the order of 10 K \cite{Mennerich08,Hay70}. For instance, in the humboldtine in which the Fe octahedron is similarly distorted \cite{Echigo08}, Ising-like behavior is observed up to 30 K \cite{Simizu88} with a similar moment orientation \cite{Sledzinska86}. The Fe octahedral symmetries in our compounds can favor a moment orientation along the twofold axes crossing the Fe$^{\rm II}$ positions, resulting in a different easy axis of magnetization for each Fe$^{\rm II}$ of the triangle at 120$^{\circ}$ from each other, as observed. The in-plane magnetic arrangement can thus reasonably be thought to be produced by $J_1$ and by a strong multi-axis anisotropy in the crystal. The 3-dimensional magnetic order is finally stabilized via interplane interactions, that can be much weaker than $T_N$. Dipolar interactions were calculated and shown to favor the observed magnetic structure. The estimated resulting dipolar energy is $E_{dip} \approx 0.18$ K, which is small compared to the anisotropy but could well induce the magnetic order through an effective antiferromagnetic interplane interaction. Thus $J_1$, a weak interplane coupling ($J_3$ or dipolar interaction) and a strong multiaxial anisotropy are sufficient to account for the ground state of these Fe-oxalates. 
This was confirmed by calculations using the discussed model (multi-axis anisotropy, NN and interplane exchange energies of  10 K, 3 K and 0.3 K respectively) that reproduced the observed magnetic structure and the magnetization curve versus field (see next section) \cite{Lhotel11}. 

\section{Magnetic properties in the ordered state}
\label{sec_LT}
\subsection{Magnetization}

\begin{figure}
\begin{center}
\includegraphics[width=7cm]{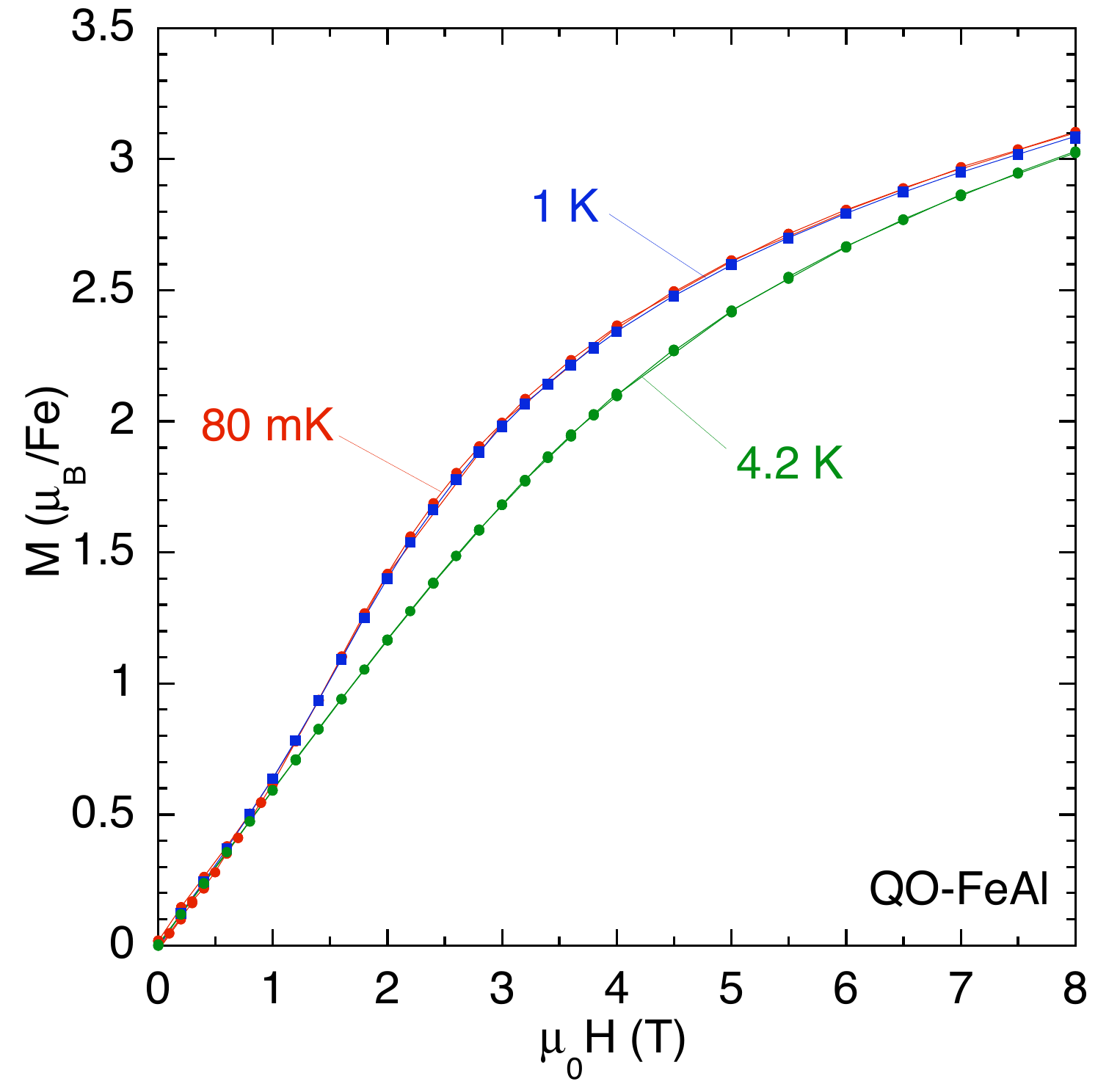}
\end{center}
\caption{Isothermals $M$ vs $H$ below 4.2 K for QO-FeAl.} 
\label{fig_FeAl_MH}
\end{figure}

\begin{figure}
\begin{center}
\includegraphics[width=7cm]{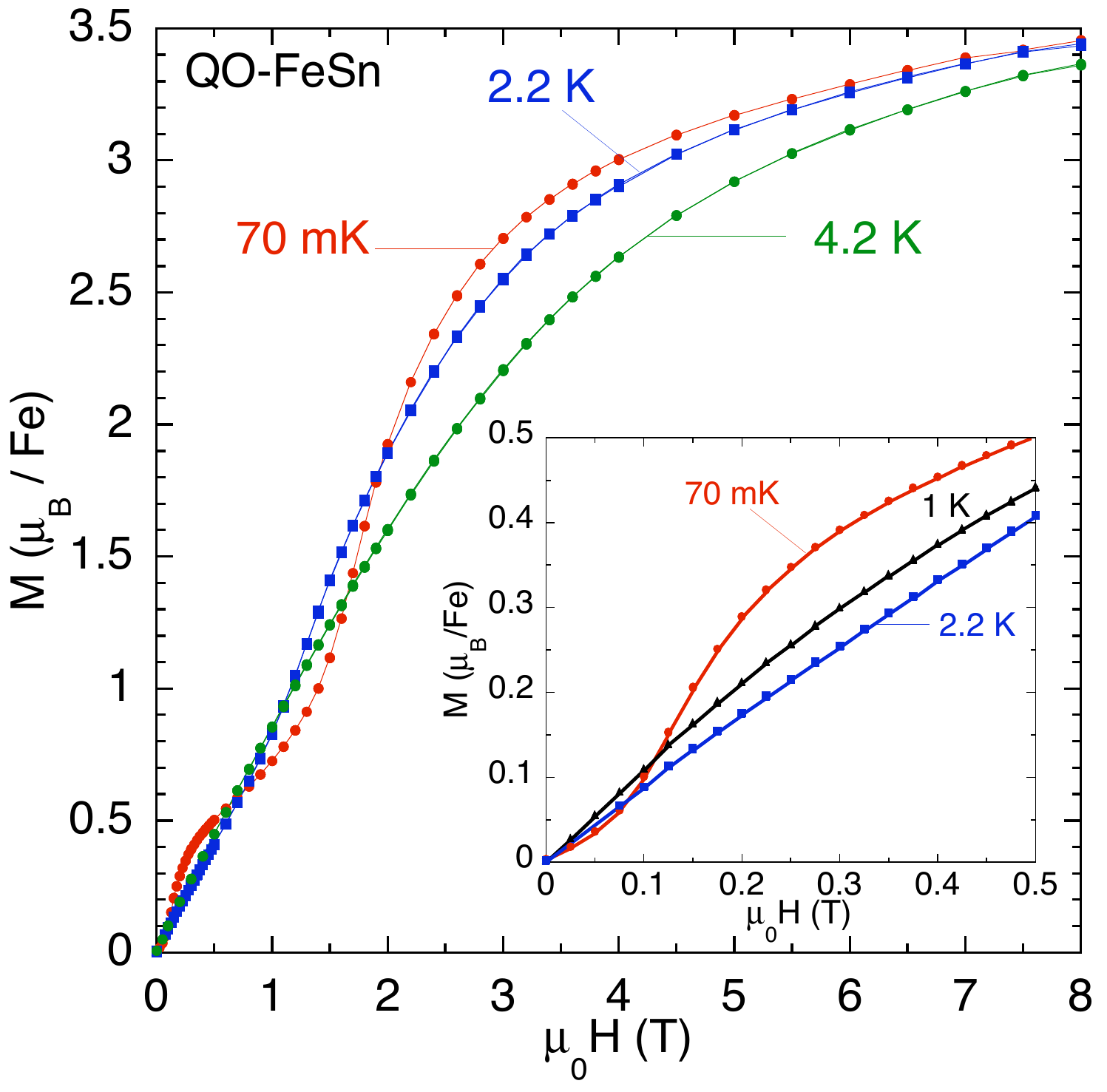}
\end{center}
\caption{Isothermals $M$ vs $H$ below 4.2 K for QO-FeSn. Inset: Zoom for temperatures below 2.2 K. } 
\label{fig_FeSn_MH}
\end{figure}

Below the N\'eel temperature, the isothermal magnetization curves show a metamagnetic like behavior around 1.7 T in all compounds (see Fig. \ref{fig_FeAl_MH} and \ref{fig_FeSn_MH}). This feature is also observed in the QO-FeAl compound, in spite of its lower $T_N$. Its magnetization value reached at 8 T at low temperature is however slightly smaller than in the other compounds (see Table \ref{table} and Fig. \ref{fig_FeAl_MH}). This might indicate a larger anisotropy in QO-FeAl. 

Mean-field calculations at zero temperature were performed considering a magnetic cell doubled along the $c$ axis, assuming that the propagation vector remains {\bf q=0} in the ({\bf a},{\bf b}) plane under magnetic field, and simulating a powder average. A model with the following hamiltonian 
\begin{eqnarray*}
{\cal H}&= &-J_1 \sum_{i,j} {\bf S_i \cdot S_j} - J_3 \sum_{i,j} {\bf S_i \cdot S_j} - D \sum_i S_i^{z_i2} \\
  & & -  \sum_i \bf{H \cdot S_i}
\end{eqnarray*}  
was adopted, where $J_1$ is the in-plane nearest-neighbor interaction, $J_3$ is the out-of-plane interaction and $D$ is the strength of  the magneto-crystalline anisotropy with respect to the local two-fold axis 
\footnote{$S_i^{z_i}$ is the projection of the spin vector on the two-fold local axis $z_i$. Within this approach, the in-plane $J_2$ exchange interaction is absorbed in an effective NN exchange interaction.}. 
With  $J_1 = -3$~K, $J_3 = 0.1J_1$ and $D=10$ K, together with $|{\bf S_i}|=1$ (for classical spins), this yields the observed magnetic structure and reproduces qualitatively the main features observed in the magnetization curves just below $T_N$ : the observed metamagnetic process and the non saturation at 8 T \cite{Lhotel11}.

\begin{figure}
\begin{center}
\includegraphics[width=7.5cm]{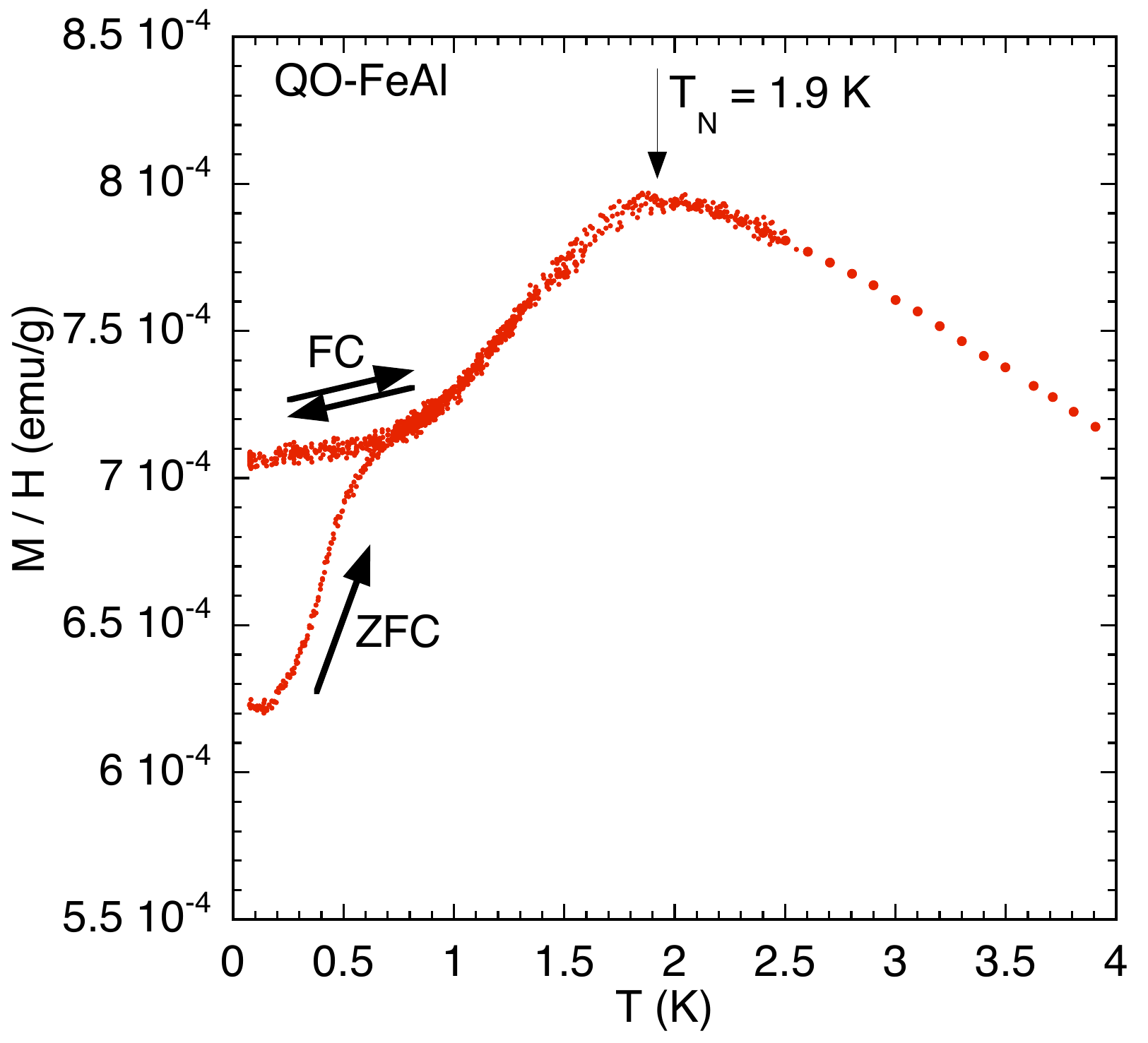}
\end{center}
\caption{$M/H$ vs T for QO-FeAl in a ZFC-FC procedure: the sample is first cooled down in zero magnetic field before the ZFC magnetization is measured under $H=500$ Oe for increasing temperature. FC Magnetization measurements are performed for decreasing then increasing temperatures in the same field of $H=500$ Oe. } 
\label{fig_FeAl_MT}
\end{figure}

Below 2 K, the magnetization measured as a function of temperature shows an upturn, and then a maximum of magnetization around 400 mK, in all the compounds except in QO-FeAl that will be discussed later on (see Fig. \ref{fig_ox_MT} and \ref{fig_FeAl_MT}). In the isothermal magnetization curves, a field-induced magnetization appears at low field below 2 K (see Fig. \ref{fig_FeSn_MH}) and a metamagnetic like behavior is observed around 0.1 T below the maximum in $M$ vs $T$ at 400 mK (see inset of Fig. \ref{fig_FeSn_MH}). Note that measurements in a Zero Field Cooled - Field Cooled (ZFC-FC) procedure show almost no irreversibility below this maximum. These low temperature features occurring in the ordered phase seem not to be associated with a change of the magnetic structure as confirmed by neutron diffraction measurements down to 60 mK in the QO-FeZr compound \cite{Lhotel11}. On the other hand, a dynamic behavior in AC susceptibility measurements is observed concomittantly, as detailed in Subsection \ref{Xac}. 

The QO-FeAl compound presents a very different behavior (see Fig. \ref{fig_FeAl_MT}). Instead of a maximum of $M$ vs $T$ below $T_N$, an irreversibility is observed in the ZFC-FC magnetization around 0.6 K, which indicates a freezing. Nevertheless, this irreversibility concerns only 10 \% of the magnetization, and no strong hysteresis is noticed in the isothermal magnetization curves, neither any frequency dependence of the AC susceptibility in this temperature range. This behavior will be discussed in Section \ref{discussion}.

\subsection{AC susceptibility}
\label{Xac}

\begin{figure}
\begin{center}
\includegraphics[width=7.5cm]{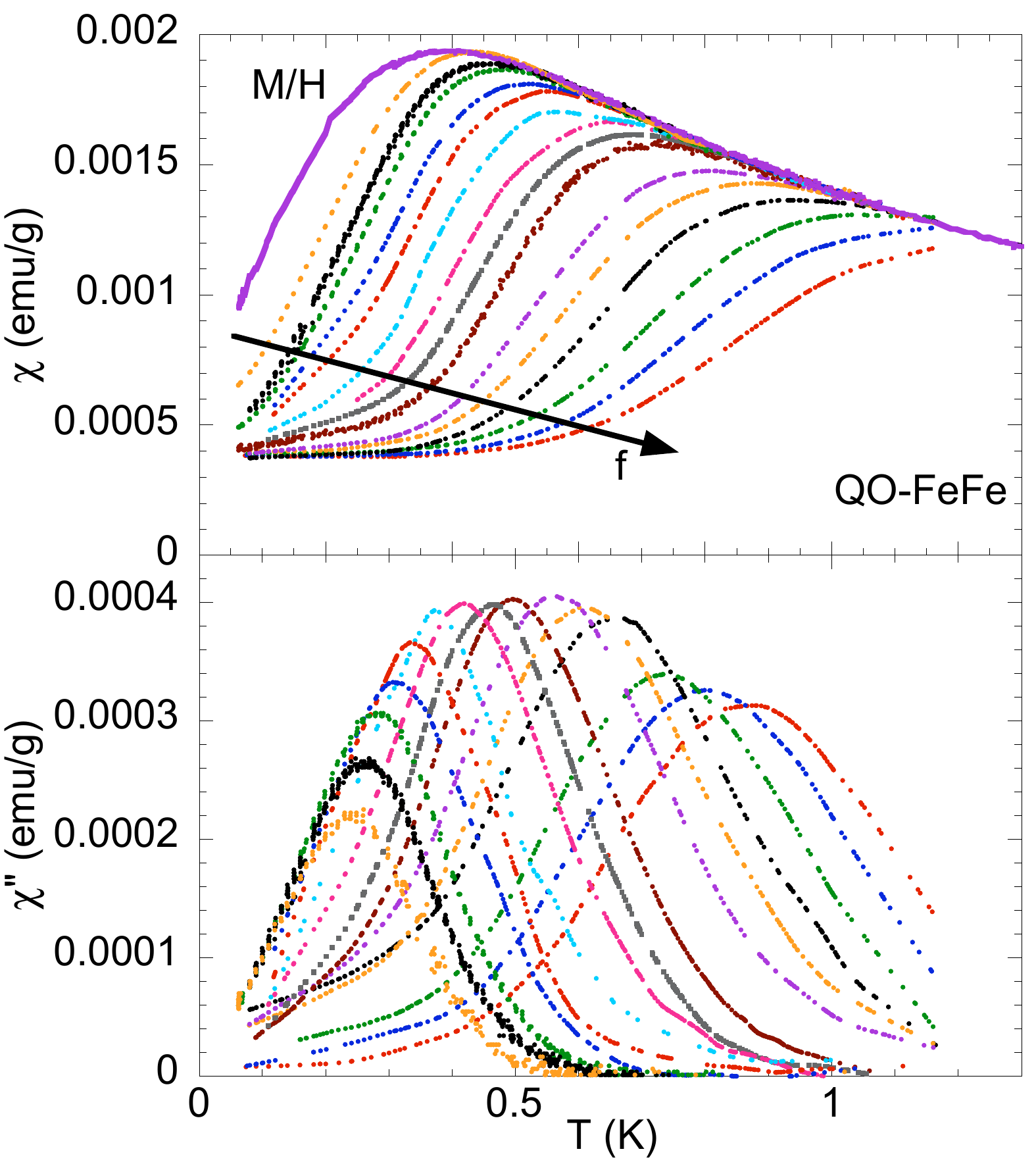}
\end{center}
\caption{AC susceptibility as a function of temperature for QO-FeFe for frequencies between 5.7 mHz and 211 Hz, with $H_{AC}=1$ Oe. Top: Real part $ \chi'$ (symbols) together with the $M/H$ curve (pink solid line). Bottom: Imaginary part $\chi''$. } 
\label{fig_FeFe_XT}
\end{figure}

\begin{figure}
\begin{center}
\includegraphics[width=7cm]{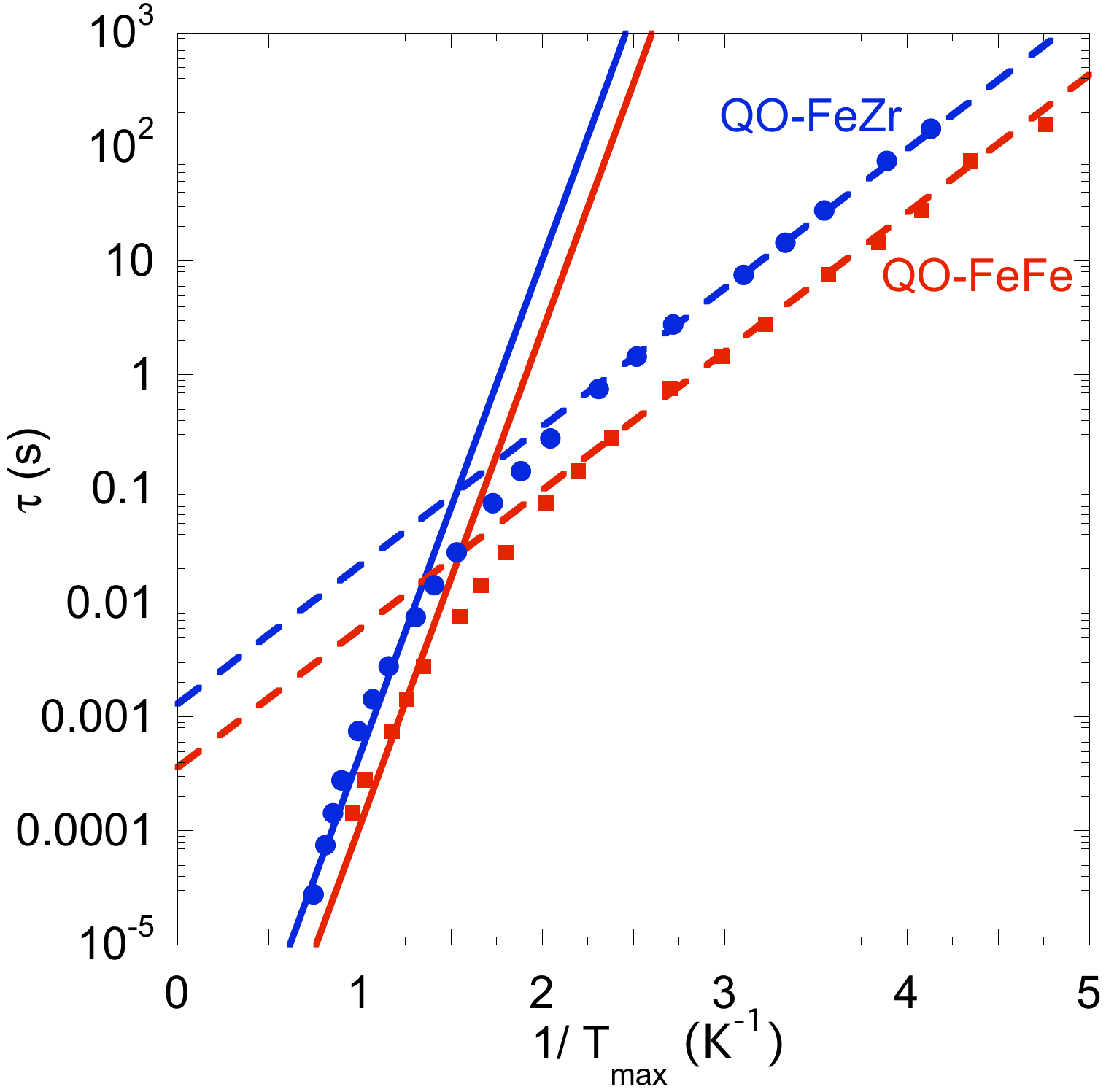}
\end{center}
\caption{$\tau$ vs $1/T_{\rm max}$ in a semilogarithmic scale for QO-FeA with A= Zr, Fe. The lines are fits to the Arrhenius law with $E_1/k_B=10$ K,  $\tau_{\rm 01-Zr}=2.1\times 10^{-8}$ s and $\tau_{\rm 01-Fe}=5\times 10^{-9}$ s (full lines)  and $E_2/k_B=2.8$ K, $\tau_{\rm 02-Zr}=1.1\times 10^{-3}$ s and $\tau_{\rm 02-Fe}=3.6\times 10^{-4}$ s (dashed lines). }
\label{fig_tau}
\end{figure}

In the three other compounds (QO-FeA with A=Zr, Sn, Fe), a frequency dependent signal was measured in AC susceptibility, as shown in figure \ref{fig_FeFe_XT} for QO-FeFe and in figure 2 of reference \cite{Lhotel11} for QO-FeZr. To probe this dependence, we have extracted the temperature $T_{\rm max}$ of the peak of the dissipative part of the susceptibility $\chi''$ for two compounds, QO-FeZr and QO-FeFe. In presence of a single relaxation time, this maximum occurs when the measurement time $\tau=1 / 2 \pi f$ (where $f$ is the frequency of the alternative field) is equal to the relaxation time of the system. We could then plot $\tau$ as a function of $1/T_{\rm max}$. The results are shown in figure \ref{fig_tau} for the two compounds. In a usual thermal activated process over an energy barrier $E$, $\tau$ would follow an Arrhenius law 
\begin{equation}
\label{eq_Arrhenius}
\tau=\tau_0 \exp(E/k_BT)
\end{equation}
where $\tau_0$ is the characteristic relaxation time. 

Figure \ref{fig_tau} shows that we have to consider two distinct regimes. The high temperature regime, above 0.8 K, is characterized by an energy barrier $E_1/k_B=10$ K and characteristic times $\tau_{\rm 01}$ of $10^{-8}-10^{-9}$ s (full lines in Fig.  \ref{fig_tau}). The low temperature regime is well described by an energy barrier $E_2/k_B=2.8$ K and $\tau_{\rm 02}=10^{-3}-10^{-4}$ s (dashed lines in Fig.  \ref{fig_tau}). The characteristic energies are taken identical in both compounds. The same value of $E_2$ is obtained from the fit in the low temperature regime for both compounds. The $E_1$ value extracted from the fit for the QO-FeZr in the high temperature regime is used to calculate the Arrhenius law for the QO-FeFe where fewer experimental points could be exploited. The intrinsic relaxation times seem to be a bit slower in the QO-FeZr compound. 

To get a deeper insight into these dynamics, we have measured the AC susceptibility as a function of frequency at a constant temperature in QO-FeZr. In the presence of a single relaxation time $\tau_m$, the complex susceptibility is given by the Casimir-du Pr\'e equation: 
\begin{equation}
\chi(\omega)=\chi'-i\chi''=\chi_S + \dfrac{\chi_T-\chi_S}{1+i\omega \tau_m}
\end{equation}
where $\omega=2\pi f$, $\chi_S$ is the adiabatic susceptibility and $\chi_T$ is the isothermal susceptibility. 

In that case, the plot of $\chi''(\omega)$ vs $\chi'(\omega)$, also called a Cole-Cole plot, would be semicircular. In the context of spin-glasses, it was proposed to introduce a phenomenological parameter $\alpha$ to account for a distribution of relaxation times in the system centered on $\tau_m$ {\cite{Huser86,Dekker89}. The resulting equation is: 
\begin{equation}
\label{eq_tau}
\chi(\omega)=\chi'-i\chi''=\chi_S+\dfrac{\chi_T-\chi_S}{1+(i\omega \tau_m)^{1-\alpha}}
\end{equation}
In Cole-Cole plots, the curve is now a flattened semicircle. The expressions of  $\chi'$ and $\chi''$ as a function of the frequency can then be deduced from this equation and compared to the experiment. 
The result at 400 mK is shown in figure  \ref{fig_sweepf}. The lines are a fit of $\chi'$ and $\chi''$ with $\tau_m=1.06$ s, $\alpha=0.38$, $\chi_S=1.76 \times 10^{-4}$ emu.g$^{-1}$ and $\chi_T=7.38 \times 10^{-4}$ emu.g$^{-1}$. The corresponding Cole-Cole plot is shown in the inset of figure  \ref{fig_sweepf}, and compared with the semicircle which would be obtained with the same parameters in absence of a distribution of relaxation times ($\alpha=0$). This analysis has been made between 200 and 900 mK, {\it i.e.} in the low temperature regime and slightly above. We recover the same temperature dependence of $\tau_m$ as obtained from the $\chi''(T)$ analysis. The $\alpha$ parameter, which is characteristic of the width of the relaxation times distribution, increases continuously on decreasing the temperature, from a value 0.2 at 900 mK to 0.5 at 200 mK. When considering an Arrhenius law (Equation \ref{eq_Arrhenius}), the resulting energy distribution $g(E)$ can be written \cite{Dekker89,LhotelPhD}:
\begin{equation}
\label{gE}
g(E)=\frac{1}{2\pi} \frac{\sin \alpha \pi}{\cosh[(1-\alpha)\frac{E-E_m}{k_BT}]-\cos \alpha \pi}
\end{equation}
Taking $E_m/k_B=2.8$ K from the low temperature regime (see Fig. \ref{fig_tau}), the half width at half maximum of the energy distribution is about 1 K and roughly constant over the investigated temperature range. This almost temperature independent energy distribution simply results from the Arrhenius expression of the energy which is proportional to the logarithm of the relaxation time multiplied by the temperature.
 
\begin{figure}
\begin{center}
\includegraphics[width=7.5cm]{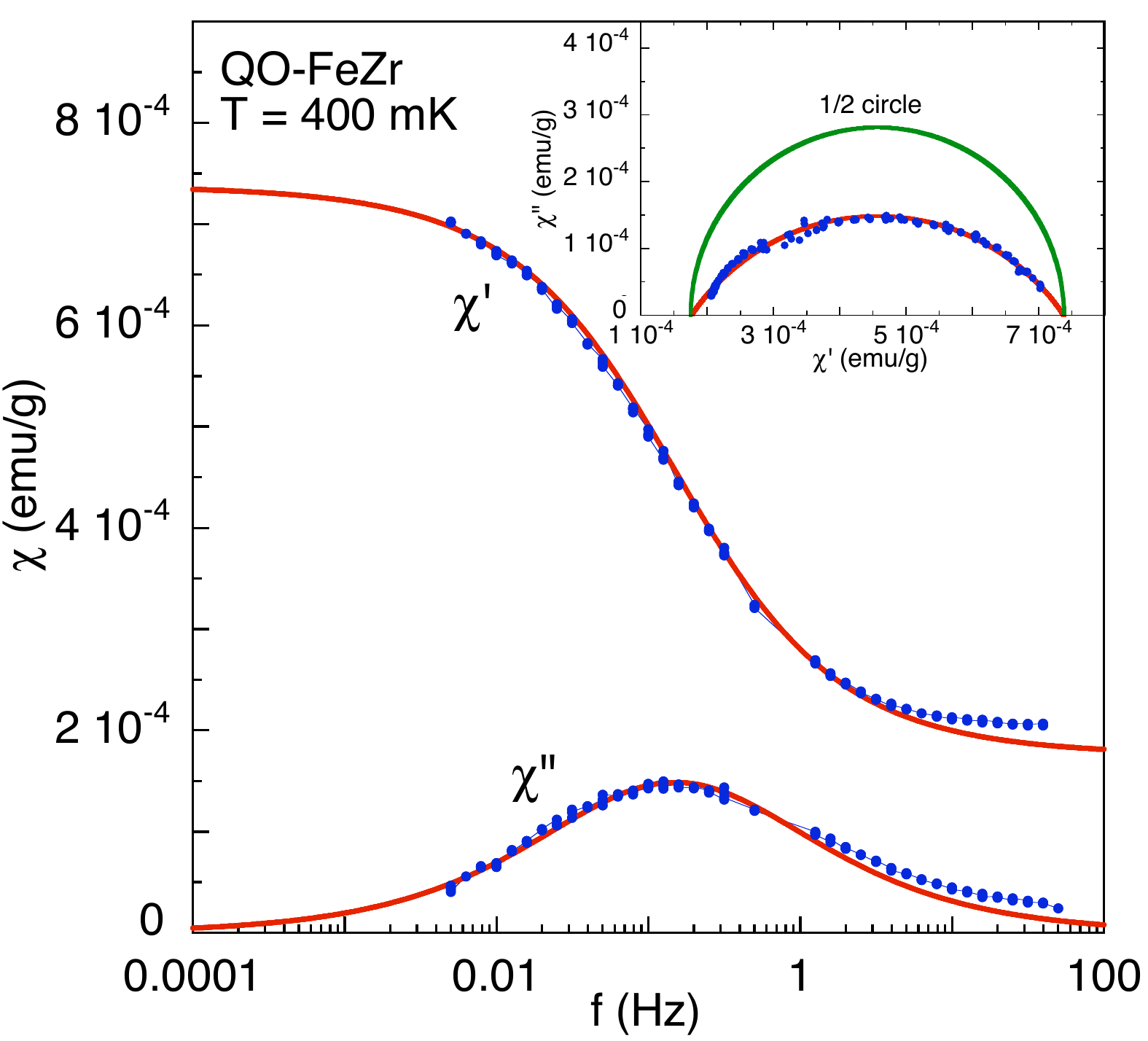}
\end{center}
\caption{AC susceptibility as a function of frequency for QO-FeZr at 400 mK. The lines are fits assuming a distribution of relaxation times following Equation \ref{eq_tau} with $\chi_S=1.76 \times 10^{-4}$ emu.g$^{-1}$, $\chi_T=7.38 \times 10^{-4}$ emu.g$^{-1}$, $\tau_m=1.06$ s and $\alpha=0.38$. Inset: Cole-Cole plot $\chi''$ vs $\chi'$. The red line is a fit with the same parameters. The green line shows the expected curve in the presence of a single relaxation time ($\alpha$=0). } 
\label{fig_sweepf}
\end{figure}

\section{Discussion}\label{discussion}

Quinternary oxalates QO-FeA compounds are the first examples of kagome-like antiferromagnets with a strong multiaxial anisotropy (estimated to 10 K), compared to the strongest exchange energies involved in the system (of the order of 3 K). This anisotropy lifts the massive degeneracy of the ground state of the kagome lattice with antiferromagnetic NN interactions and results in an in-plane {\bf q=0} magnetic structure as observed in the QO-FeA. 

Although the frustration is released, the magnetic properties of the QO-FeA (A=Zr, Sn, Fe) compounds below $T_N$, in particular the existence of slow spin dynamics in the ordered magnetic phase, are radically different from usual antiferromagnets. We have shown that these features are directly related to the low connectivity of the kagome lattice \cite{Lhotel11}.  Indeed, two types of domains where all the spins are reversed in the kagome planes are allowed by symmetry in this magnetic structure. Due to the topology of the kagome lattice, a spin at the boundary between two such domains is only connected to two triangles, each one belonging to one domain (see Fig. \ref{fig_domain}). More important, this boundary spin is not connected to the other spins of the domain wall. The energy resulting from the interaction of  such a boundary spin with the other spins is therefore the same in any of its two orientations compatible with the easy axis of magnetization. This spin is then free to flip above the anisotropy barrier. This is actually the case if the NN exchange interaction $J_1$ is considered and remains true in presence of the NNN exchange interaction $J_2$, provided that the boundary spin is not at a wall corner. The "high" temperature dynamics observed in AC susceptibility (see Section \ref{Xac}) above 0.8 K can be ascribed to this phenomenon, the energy barrier $E_1=10$ K being the anisotropy barrier, and the characteristic time $\tau_{01}$ being the intrinsic time for a single spin-flip. This interpretation in terms of exchange-released spins is consistent with the absence of hysteresis loops, since the dynamics are local and do not result in a motion of the domain-walls. 

A parallel can be attempted with the slow dynamics observed in weakly coupled single-chain magnets with a strong anisotropy \cite{Coulon2006,Okuda81,Lhotel06}. In these systems a single spin can flip above the anisotropy barrier at a point domain wall along a chain, again due to the low connectivity, inherent this time to the quasi-one dimensionality of the system. However, in that case, the spin reversal can propagate along the spin-chain, thus inducing hysteresis loops. 

When decreasing the temperature, we enter a second regime below 0.8 K, in which the spin dynamics are governed by a much longer characteristic time $\tau_{02}$ of about 10$^{-3}$-10$^{-4}$ s which indicates a change in the spin flip mechanism. In addition, the apparent energy barrier is strongly reduced: $E_2=2.8$ K. It is very similar to the change of regime of the slow spin dynamics that has been observed around 12 K in the dipolar spin ices \cite{Matsuhira01,Snyder04} where it was interpreted as the onset of quantum tunneling and its influence on the single spin flips regime (thermally activated above an anisotropy barrier at higher temperature). In the present compounds, the quantum tunelling could explain the partial erasing of the energy barrier. In addition, this low temperature regime occurs at temperatures where further interactions are expected to play a role, in particular long-range dipolar interactions which couple the exchange released spins along the domain wall. The large distribution width of the energy barrier measured at low temperature might be induced by the dipolar energy distribution. 

We have already underlined the similarities between the observed spin dynamics in the QO-FeA ordered phase and the slow spin dynamics observed in dipolar spin ices. The analogy can be pursued on a larger nanometric scale with sea ice: in a porous solid ice matrix, liquid inclusions flow in the interstices, and get amorphously frozen as the temperature is decreased \cite{Wettlaufer97}. The topology of the kagome lattice provides a medium similarly sustaining free spins at the boundary between the antiferromagnetic domains, before they become correlated through dipolar interactions. 

\begin{figure}
\begin{center}
\includegraphics[width=8cm]{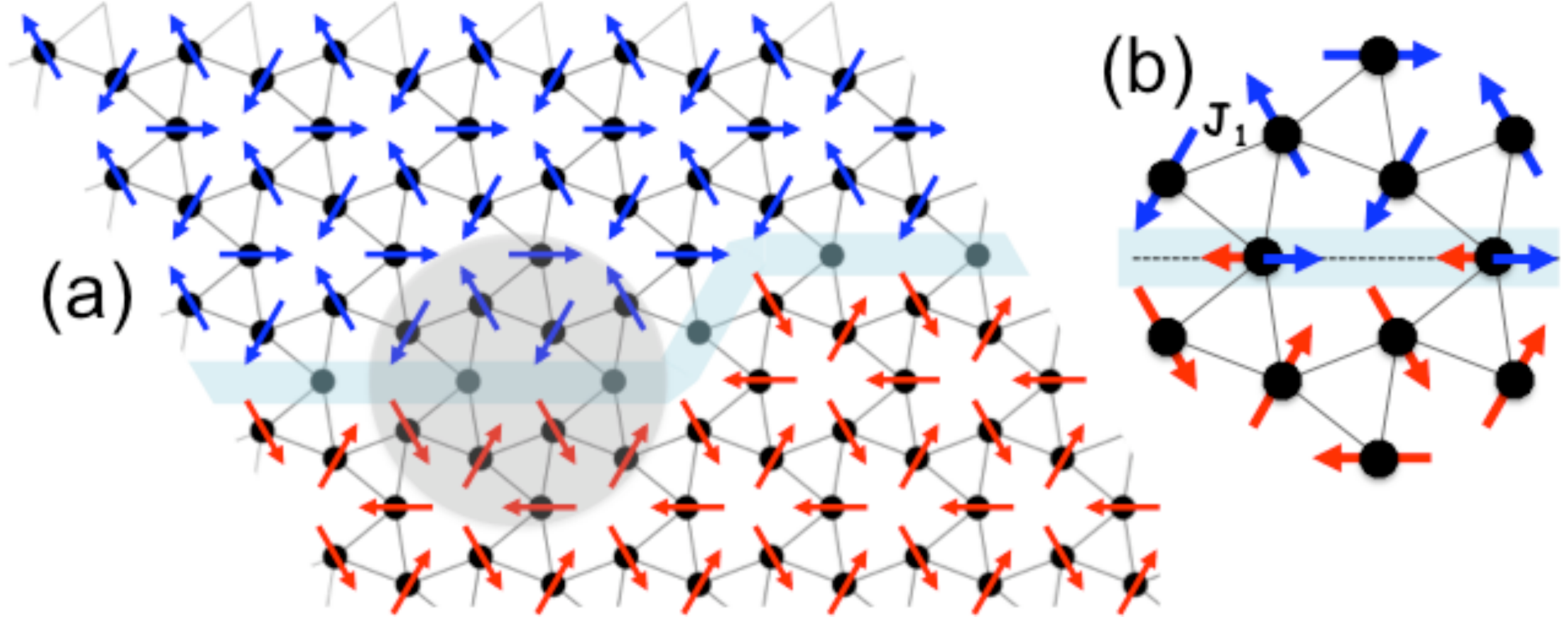}
\end{center}
\caption{(a) Two antiferromagnetic domains in the QO-FeA distorted kagome lattice. The blue stripe underlines the boundary exchange released spins. (b) Zoom showing two boundary spins that can point in two energetically equivalent directions compatible with the easy axis (blue and red arrows).  } 
\label{fig_domain}
\end{figure}

Finally, an unexpected result in this study is the strong difference of the magnetic properties of the QO-FeAl compound, compared to the other members of the family. This is even more amazing since the presence of Fe$^{\rm IIl}$ paramagnetic spins in the QO-FeFe compound does not affect its magnetic properties. Although the structures of the four compounds are very similar as shown by neutron diffraction, only QO-FeAl has a smaller N\'eel temperature and a smaller saturation value of the magnetization (see Table \ref{table}). These results point respectively toward a smaller NN exchange interaction $J_1$ and a larger magneto-crystalline anisotropy. Although the reason for these modifications remains unclear, the main difference of this compound lies in the size of the Al ion, which is much smaller than its counterparts (Zr, Sn and Fe) in the other compounds. This might modify the distribution of the orbitals, thus affecting the exchange paths and the anisotropy. 

Furthermore, instead of the dynamics observed in other compounds by AC susceptibility, QO-FeAl rather presents a small freezing of the magnetization below 600 mK. In the presence of a larger anisotropy, the reversal time of a single spin in the boundary might be too long to be probed by our AC susceptibility measurements. In our experimental time scale, the boundary spins would then appear frozen, thus inducing a small ZFC-FC irreversibility. 

\section{Conclusion}

To summarize, we have reported the study of a new family of metallo-organic materials with Fe$^{\rm II}$ ions and oxalate ligands. These QO-FeA compounds have enabled the investigation of a magnetic (distorted) kagome lattice with exchange and dipolar interactions, and multiaxial anisotropy. The latter induces a noncollinear magnetic order, which sustains  slow spin dynamics ascribed to the spins located at the boundary between the antiferromagnetic domains. The observed spin dynamics could be generic to geometrically frustrated magnets with a low lattice connectivity, where residual spin fluctuations are often reported in the ordered phase \cite{Mirebeau08}. 
\bigskip

We would like to thank P. Convert for his help during the first neutron experiment on D20. DJP, PTW and AKP are grateful for financial support from the EPSRC.   

%

\end{document}